\documentclass[11pt,a4paper]{article}
\usepackage{jheppub}

  \usepackage{amsmath}
  \usepackage{amsfonts}
  \usepackage{amssymb}

  \usepackage{graphicx}
  \usepackage{float}
  \DeclareGraphicsExtensions{.ps}

  \newcommand{\be}{\begin{equation*}}
  \newcommand{\ee}{\end{equation*}}
  \newcommand{\ben}{\begin{equation}}
  \newcommand{\een}{\end{equation}}
  \newcommand{\bse}{\begin{subequations*}}
  \newcommand{\ese}{\end{subequations*}}
  \newcommand{\bsen}{\begin{subequations}}
  \newcommand{\esen}{\end{subequations}}
  
    \newcommand{\full}{{\phi}}
    \newcommand{\qfull}{\hat{\phi}}
    \newcommand{\mean}{\Phi}
    \newcommand{\pert}{\delta\hat{\phi}}
    \newcommand{\ensemble}{\psi_i}
    \newcommand{\mode}{f_k}
    \newcommand{\AO}{A_0}
    \newcommand{\rO}{r_0}
    \newcommand{\xt}{\mathbf{x},t}
    \newcommand{\xz}{\mathbf{x},0}
    \newcommand{\intxxt}{\int d^2x\,dt}
    \newcommand{\ak}{a_k}
    \newcommand{\akt}{a_k^{\dagger}}
    \newcommand{\intkk}{\int\frac{\mathrm{d}^2{k}}{\left(2\pi\right)^2}}
 
\title{Oscillon Lifetime in the Presence of Quantum Fluctuations}

\date{\today}

\author[a]{Paul M. Saffin,}

\author[a]{Paul Tognarelli,}

\author[b]{Anders Tranberg}

\affiliation[a]{School of Physics and Astronomy, University Park, University of Nottingham,\\ Nottingham NG7 2RD, United Kingdom}
\affiliation[b]{Faculty of Science and Technology, University of Stavanger, 4036 Stavanger, Norway}

\abstract{
We consider the stability of oscillons in 2+1 space-time dimensions, in the presence of quantum fluctuations. Taking the oscillon to be the inhomogeneous mean field of a self-interacting quantum scalar field, we compare its classical evolution to the evolution in the presence of quantum fluctuations. The evolution of these and their back reaction onto the mean field is implemented through the inhomogeneous Hartree approximation, in turn computed as a statistical ensemble of field realizations.  We find that although the lifetime of the oscillon is dramatically reduced compared to the classical limit, the regions of longevity are similar in the space of Gaussian initial configurations. 
}
\emailAdd{paul.saffin@nottingham.ac.uk}
\emailAdd{ppxpt@nottingham.ac.uk}
\emailAdd{anders.tranberg@uis.no}


\begin{document}

\maketitle

\section{Introduction}
\label{sec:intro}

Oscillons are a class of very long-lived, quasi-periodic, non-topological soliton observed to arise in various, non-linear field-theories \cite{oscillons, bi:Salmi2012, bi:Gleiser2007}.  Classically, oscillon lifetimes vary from $\mathcal{O}(10^4)$ in natural units in 3+1 space-time dimensions through to larger than $\mathcal{O}(10^6)$ in 2+1 dimensions (see for instance \cite{bi:Salmi2012, bi:Saffin2007}), and even stable solutions (such as the Sine-Gordon breather) can exist in 1+1 dimensions.  This stability is not enforced by the topology of the theory, such as for topological defects, or any conserved charges, such as for Q-balls. Instead, stability arises from the large-amplitude oscillations of the configuration being non-linear and the basic frequency smaller than the particle mass \cite{bi:Saffin2007, bi:oscillon-frequency, bi:Andersen2012}.  This makes it hard for the oscillation (at least, perturbatively) to excite particle modes of the field, triggering decay and loss of energy.

Oscillons are known to be generated in phase transitions \cite{bi:Gleiser2007, bi:Amin2010, bi:Gleiser2003}.  In such situations, the oscillons can carry a significant fraction of the energy available; the long decay time may consequently influence the thermalization time of matter after cosmological phase transitions \cite{bi:Gleiser2003}.  In the intermediate state, oscillons supply regions where the field expectation value is away from the vacuum value. It therefore departs from equilibrium and may source particle creation and baryogenesis.

Although very long-lived, over time an oscillon does slowly shed energy, eventually reaching a critical, lowest-energy  profile that collapses through a rapid, though poorly understood, non-linear process (see however \cite{bi:Saffin2007, bi:Andersen2012}).  The semi-classical scalar dynamics on the background of the classical oscillon indicate that the decay rate is increased and even dominated by the addition of quantum effects \cite{bi:Hertzberg2010}.  In progresssion from the classical regime to the quantum-field theoretic description, a pertinent question remains whether the shedding of energy is altered, potentially greatly accelerated, by the presence of quantum degrees of freedom offering additional decay-channels to the non-linear configuration.

Just as for Q-balls, in addition to decay to other particle species, one could imagine a tunnelling transition to a lower-energy state (say, of smaller oscillons or elementary excitations) [5]. Also, the shape of the effective potential may change in such a way that stability is lost for some classically stable configurations [6]. The potential results of the quantum self-interaction is very hard to compute perturbatively in a time- and space-dependent mean field background; but numerically it is possible, as we will demonstrate here.

In the subsequent sections, we setup a simple 2+1 dimensional, single-scalar field in a quartic potential and discuss classical oscillons.  We then obtain the quantum mean-field equations that are analogue to the classical field equations, and the equation of motion for the quantum perturbations to the classical system. The combined equations describe the quantum dynamics that we intend to investigate.  We introduce at that point how the quantum state can be approximated by a Gaussian ensemble of random initial realizations and set up the numerical approach for evolving the dynamical equations. We use this to examine the lifetime of the oscillons determined from the simulations and compare the region of stability and the lifetimes to their classical counterparts. We then conclude.

\section{Setup and Model}
\label{sec:SaM}

We consider a single real scalar field in $2+1$ dimensions, with the action
\be
S = -\intxxt \left[\frac{1}{2}\partial_{{\mu}}\full\,\partial^{{\mu}}\full +\frac{1}{2}m^2\full^2  + \frac{\lambda}{4}\full^4\right].
\ee
Variation of the action yields the equation of motion for the classical system
\ben
\label{eq:SaM_EoM}
\left[\partial^2_{{t}} - {\nabla}^2 + m^2 + \lambda\full^2\left(\xt\right)\right]\full\left(\xt\right) = 0\mbox{.}
\een

Setting $m^2<0$, the potential has two minima with vacuum expectation values $v^\pm = \pm\sqrt{-m^2/\lambda}$.  
This choice for the potential also allows for oscillons to exist, and a rough criterion is that the centre and some fraction of the localized oscillon configuration should have amplitude beyond the inflection point of the potential $\phi=\sqrt{-m^2/3\lambda}$. There is evidence \cite{bi:Andersen2012} that there is only a single "trajectory" of oscillons and that it is an attractor in field space: so although an analytic closed form for the oscillon profile is unknown, starting the field evolving from within the basin of attraction for this configuration is sufficient to establish an oscillon.  Taking the initial configuration to be a Gaussian over the broken phase vacuum
\ben
\label{eq:SaM_initial_condition}
\full\left(\mathbf{{x}}, 0\right)= v^+\left[1 - \AO\exp\left(-\frac{{x}^2+{y}^2}{2\rO^2}\right)\right]\mbox{,}
\een
provides a suitable profile to produce an oscillon, for certain values of the amplitude $\AO$ and width $\rO$. Within the basin of attraction, different values of these parameters lead to evolution into an oscillon at different points along the "trajectory". Once there, memory of the initial condition is lost, and the oscillon evolves in a unique way along the trajectory, until it finally collapses.

Rescaling the system according to $\phi \rightarrow \sqrt{|m^2|/\lambda}\full$ and $x \rightarrow x/m$ effectively sets the mass and coupling equal to unity in the equation of motion for the classical scalar field, and so general masses and couplings may be traded in for the field and time normalization. The quantum dynamics, however, introduces another scale, the scale of fluctuations (which is essentially $\hbar$). We will press on with the unrescaled field, but ultimately use $\lambda/m=1$ and $\hbar=1$ for our simulations, so that time is in mass units and the ratio between amplitude of quantum fluctuations and the vev is fixed by hand.

In the quantum theory, $\full$ is promoted to a field operator and we split the operator into its mean $\mean = \langle\qfull\rangle$ and a perturbation $\pert$:
\be
\qfull\left(\xt\right) = \mean\left(\xt\right) + \pert\left(\xt\right).
\ee
The evolution equations for the mean field and mode functions follow from first promoting the classical equation of motion (\ref{eq:SaM_EoM}) to an operator equation (the Heisenberg equation).  Then taking the expectation value of the operator equation \cite{bi:Hindmarsh2009,bi:Salle2001}, we find for the mean field (ignoring connected correlators beyond quadratic)
\ben
\label{eq:MF}
\left[\partial_{{t}}^2 - {\nabla}^2 - m^2 + \lambda\mean^2 + 3\lambda\langle\pert\left(\xt\right)\pert\left(\xt\right)\rangle\right]\mean\left(\xt\right) = 0,
\een
and for the perturbations
\ben
\label{eq:modes}
\left[\partial^2_{{t}} - {\nabla}^2 - m^2 + 3\lambda\mean^2 + 3\lambda\langle\pert\left(\xt\right)\pert\left(\xt\right)\rangle\right]\pert\left(\xt\right) = 0\mbox{.}
\een
For a Gaussian truncation (such as the current Hartree approximation), we can expand the perturbation into orthogonal mode functions $\mode$:
\ben
\label{eq:modes2}
\pert\left(\xt\right) = \intkk \left(\ak\mode\left(\xt\right) + \akt\mode^*\left(\xt\right)\right),
\een
where each mode function independently satisfies the perturbation equation of motion (\ref{eq:modes}).  
The $\{\ak\}$ and $\{\akt\}$ are the ladder operators, and are time-independent.  They obey
\be
\left[a_k,a^\dagger_{k'}\right] = (2\pi)^2\delta^2({\bf k-k'})\mbox{, }\langle\akt\ak\rangle = n_{{k}}(2\omega_k)(2\pi)^2\delta^2({\bf k-k'}),
\ee
where the $n_k$ correspond to the particle number in the free scalar theory. In the Hartree approximation (although more general than the free field theory case), the $n_{{k}}$ still provide a sensible definition of the occupation number in a given quantum mode.

We initialize the quantum mean field $\mean$ with the Gaussian profile (\ref{eq:SaM_initial_condition}).  In the absence of back-reaction from the perturbations in the equation (\ref{eq:MF}), this initial configuration would obey the classical dynamics and for certain values of $A_0$, $r_0$ evolve into an oscillon. When the two-point correlator of the perturbations is small, the initial configuration remains perturbatively close to the expected, oscillon solutions of the quantum mean-field dynamics.

For the mode functions $\mode$, we chose the initial configuration to be that of a translationally invariant system. In particular, in the free vacuum ($\lambda = 0$), we have the familiar plane wave solutions:
\ben
\label{eq:QP_intial-mode-functions}
\mode\left({\bf x},0\right) = \frac{1}{2\omega_{{k}}}\exp\left(i\mathbf{{k}}\cdot\mathbf{{x}}\right),\qquad \partial_t f_{k}({\bf x},0)=\frac{i}{2}\exp\left(i\mathbf{{k}}\cdot\mathbf{{x}} \right).
\een
Notably, with this choice for the initial condition, the initial two-point correlators of the perturbations become
\begin{eqnarray}
\langle\delta\hat{\phi}\left(\xz\right)\delta\hat{\phi}\left(\xz\right)\rangle = \intkk \frac{1}{\omega_{{k}}}\left[n_{{k}} + \frac{1}{2}\right],\\
\langle\delta\hat{\pi}\left(\xz\right)\delta\hat{\pi}\left(\xz\right)\rangle = \intkk\, \omega_{{k}}\left[n_{{k}} + \frac{1}{2}\right].
\end{eqnarray}
The first of these enters as the fluctuation, back-reaction contribution. It is divergent, and we perform a mass renormalization through the counterterm
\ben
m^2 \rightarrow m^2 - 3\lambda\langle\pert\left(\xz\right)\pert\left(\xz\right)\rangle,
\een
to yield a finite, physical mass\footnote{In $2+1$ dimensions only this linear divergence is present. Beyond the Hartree approximation, further divergences would be present.}.

The quantum fields can be rescaled, as the classical fields, according to $\mean \rightarrow \sqrt{|m^2|/\lambda}\mean$ and $\pert \rightarrow \sqrt{|m^2|/\lambda}\pert$, along with the rescaling of the co-ordinates $x \rightarrow x/m$. This sets the mass and coupling effectively equal to unity, in exact parallel to the classical rescaling, for both the mean and perturbation dynamics. Briefly reinstating $\hbar$, the vacuum two-point correlator becomes (remembering that $k \rightarrow mk$ and $\omega_k \rightarrow m\omega_k$)
\ben
\left\langle\delta\hat{\phi}\left({\bf x}\right)\delta\hat{\phi}\left({\bf x}\right)\right\rangle = \intkk \frac{\hbar\lambda}{m}\frac{1}{\omega_{{k}}}\left[n_{{k}} + \frac{1}{2}\right],
\een
where $\omega_k=\sqrt{2+k^2}$. As mentioned, we now take $\hbar=\lambda/m=1$, which is a particular choice of the scale of fluctuations relative to the vev.  Ultimately, we also set the occupation number $n_k = 0$.

The most straightforward way to numerically solve the system of equations (\ref{eq:MF}) and (\ref{eq:modes}) is to insert the ansatz (\ref{eq:modes2}) into (\ref{eq:modes}), thereby generating a set of $N^2+1$ coupled differential equations (where $N$ is the linear size of the spatial lattice and the `+1' accounts for the mean field equation (\ref{eq:MF})). The coupling is through the appearance of the correlator 
\ben
\label{eq:ET_modes->two-point}
\left\langle\pert\left(\xt\right)\pert\left(\xt\right)\right\rangle = \intkk \left[n_{{k}} + \frac{1}{2}\right]\left|\mode\left(\xt\right)\right|^2,
\een
in both mode and mean field equations. The main difficulty with this technique is the very large number of mode functions, since in a time-dependent, inhomogeneous background the $f_{k}({\bf x},t)$ are each space-dependent fields, and so the numerical problem scales as $N^4$.   

The alternative method is to replace the computation (\ref{eq:ET_modes->two-point}) with an average over a new ensemble of fields $\psi_i$:
\ben
\label{eq:replacement}
\left\langle\delta\hat{\phi}\left(\xt\right)\delta\hat{\phi}\left(\xt\right)\right\rangle \rightarrow \left\langle\ensemble\left(\xt\right)\ensemble\left(\xt\right)\right\rangle_E.
\een
Each of these new field realizations is evolved according to (\ref{eq:modes}), i.e. in position space, and is constructed by taking the random complex-numbers $\{c_i^k\}$, and writing at the initial time
\ben
\psi_i({\bf {x}},0)=\sum \frac{d^2 {k}}{(2\pi)^2}\frac{1}{2\omega_k} \left[c_i^k e^{i{\bf kx} }+(c_i^k)^*e^{-i{\bf k x}}\right].
\een
Then for each $k$, the $\{c_i^k\}$ are Gaussianly distributed with zero mean and
\ben
\label{eq:ET_<random>-equals-<operator>}
\left\langle c_i^k(c_i^{k'})^*\right\rangle_E = \left[n_k + \frac{1}{2}\right](2\omega_k)(2\pi)^2\delta({\bf k}-{\bf k'})
\een
so that the linearity of the equation of motion ensures that the replacement (\ref{eq:replacement}) is correct in the limit of many random realizations, $i=1,...,M$ for $M$ large. This implementation scales as $N^2\times M$, which is potentially smaller than $N^4$ (although we will see here that the gain for inhomogeneous systems of the present type is only a factor of a few).

\section{Classical and quantum oscillons}
\label{sec:oscillon}

With the system discretized on a spatial lattice of spacing $\delta x=0.8m$, we further discretized the time-evolution with a time step $\delta t=0.05m$. For this highly inhomogeneous system, we found that a very large number of realizations $M$ was necessary, with convergence being reached only around $M=60000$. For a lattice of $N=256$, this is a very small gain in computer time, compared to solving all the mode equations. This is in contrast to the work of \cite{bi:Borsanyi2008} and \cite{bi:Hindmarsh2009} that demonstrated much better convergence, however for volume averaged quantities; but is consistent with \cite{weir} and \cite{fermions} for respectively, bosons and fermions in truly inhomogeneous systems. 

We scanned the space of Gaussian initial profiles in the region $A_0\in \left[-4, 4\right]$ and $r_0\in \left[0, 4\right]$, inspired by the results of \cite{bi:Andersen2012}. This nicely covers the main boundaries of the classical oscillon basin of attraction. In most cases, identifying the decay of the oscillon is straightforward but especially near the region's edge, it becomes harder. For the quantum simulations, we argue from the structure of the classical parameter-space that a reduced region $A_0\in \left[0.25, 4\right]$ and $r_0\in \left[1, 4\right]$ is sufficient to cover the full corresponding-region of interest while reducing the required computing time. 

A square-grid of $N=256$ with the converged ensemble-size $M=256^2$ was used for each set $(A_0,r_0)$ to provide computations that were feasible within a reasonable duration. Selected parameter sets were repeated on larger grids of $N=384$ and $N=512$ (also with larger $M$).  These enabled us both when the lattice spacing was fixed, to check for finite volume effects and to test for resolution effects when maintaining a constant physical volume.

\begin{figure}[H]
  \centering
    \includegraphics[width = 0.49\textwidth]{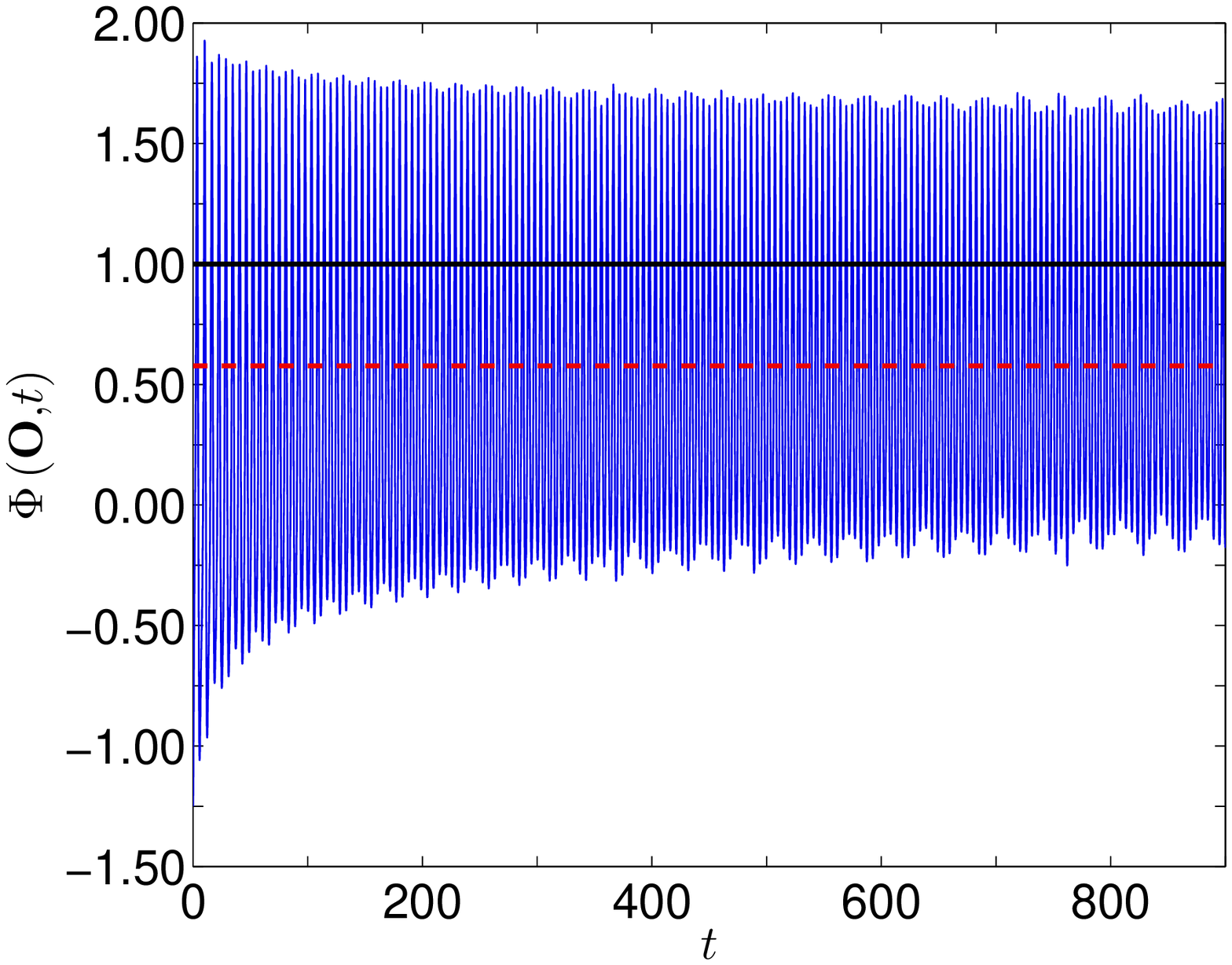}
    \includegraphics[width = 0.49\textwidth]{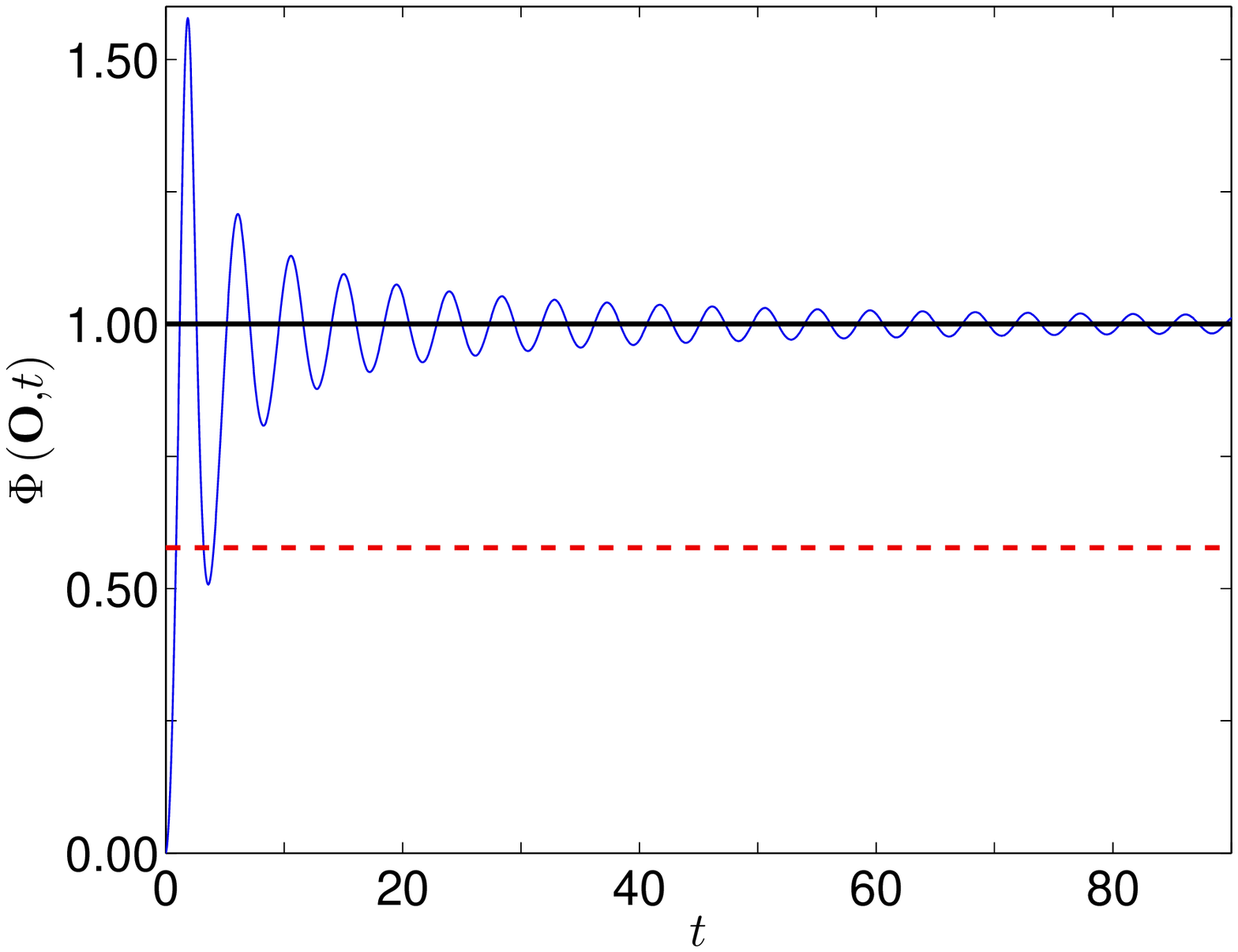}
    \includegraphics[width = 0.49\textwidth]{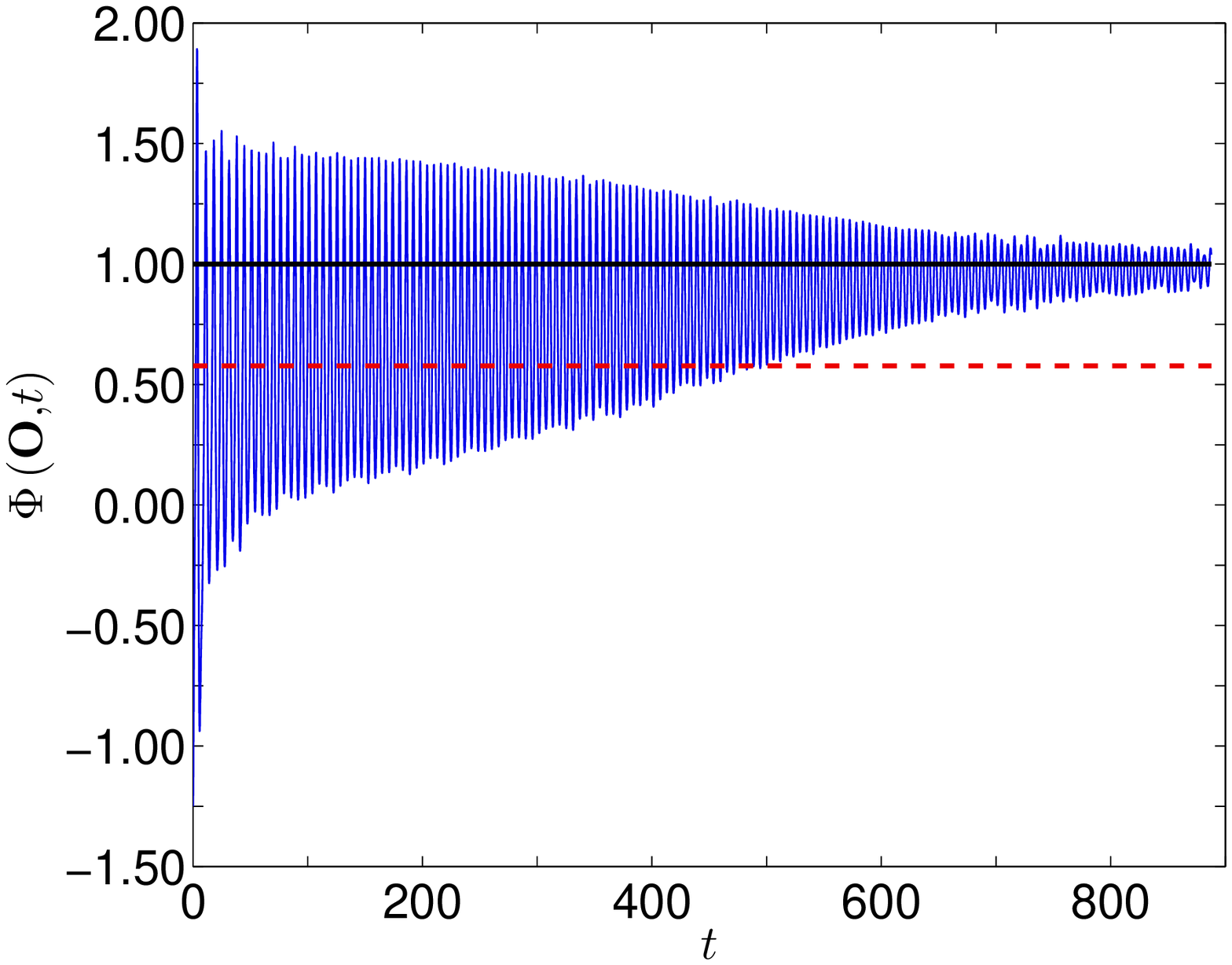}
    \includegraphics[width = 0.49\textwidth]{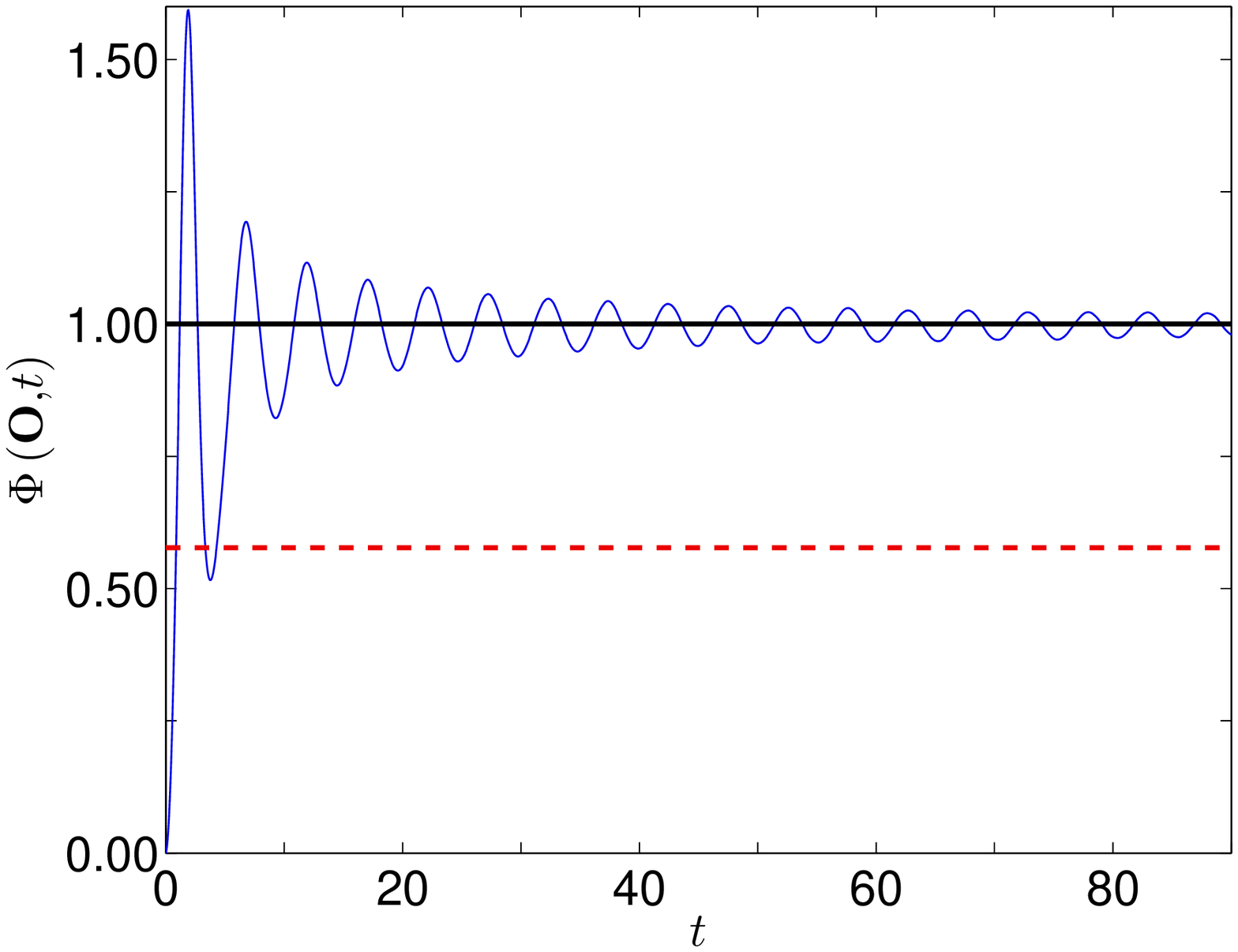}
  \caption{The time evolution of the center of a configuration with initial condition inside (left) and one outside (right) the basin of attraction for the classical (top) and quantum (bottom) oscillon. $(A_0,r_0)=(2.25, 2.5)$ and $(1.0, 1.5)$, respectively. }
  \label{fig:liveanddie}
\end{figure}

Fig. \ref{fig:liveanddie} shows two different initial conditions, inside and outside the basin of attraction respectively. In the two top plots, these are evolved classically, and we see that inside the basin of attraction, the oscillon settles and then lives on for a much longer time than the simulation. Outside, the centre amplitude decays below the inflection point in two periods or less. 
In the quantum evolution (the two bottom plots), we again see that outside the basin of attraction, the evolution is very similar: an oscillon is never generated, and the quantum back-reaction never has any impact. Inside the basin, however, the evolution is very different. The oscillon clearly emerges, though with a somewhat narrower oscillation envelope. Over roughly 100 periods of oscillation, however, the amplitude gradually decreases to a more or less harmonic oscillation no longer crossing the point of inflection for the classical potential. 

This suggests a number of things about the quantum system. There is an effective quantum potential, which does allow for long-lived oscillon-like solutions. Some Gaussian initial profiles are again close enough to such a solution to be attracted by it. Lifetimes are likely much shorter than in the classical case, and the decay process is very different. Whereas for classical oscillons, decay takes place as a sudden collapse, for the quantum case, the decay and loss of energy is gradual throughout. In particular, nothing dramatic happens near the classical inflection point, suggesting that this point has no connection to the shape of the effective potential. Interestingly, the minimum of the potential agrees with the classical case, and so our renormalization procedure works well, even though it is approximate. 

\begin{figure}[H]
  \centering
    \includegraphics[width = 0.49\textwidth]{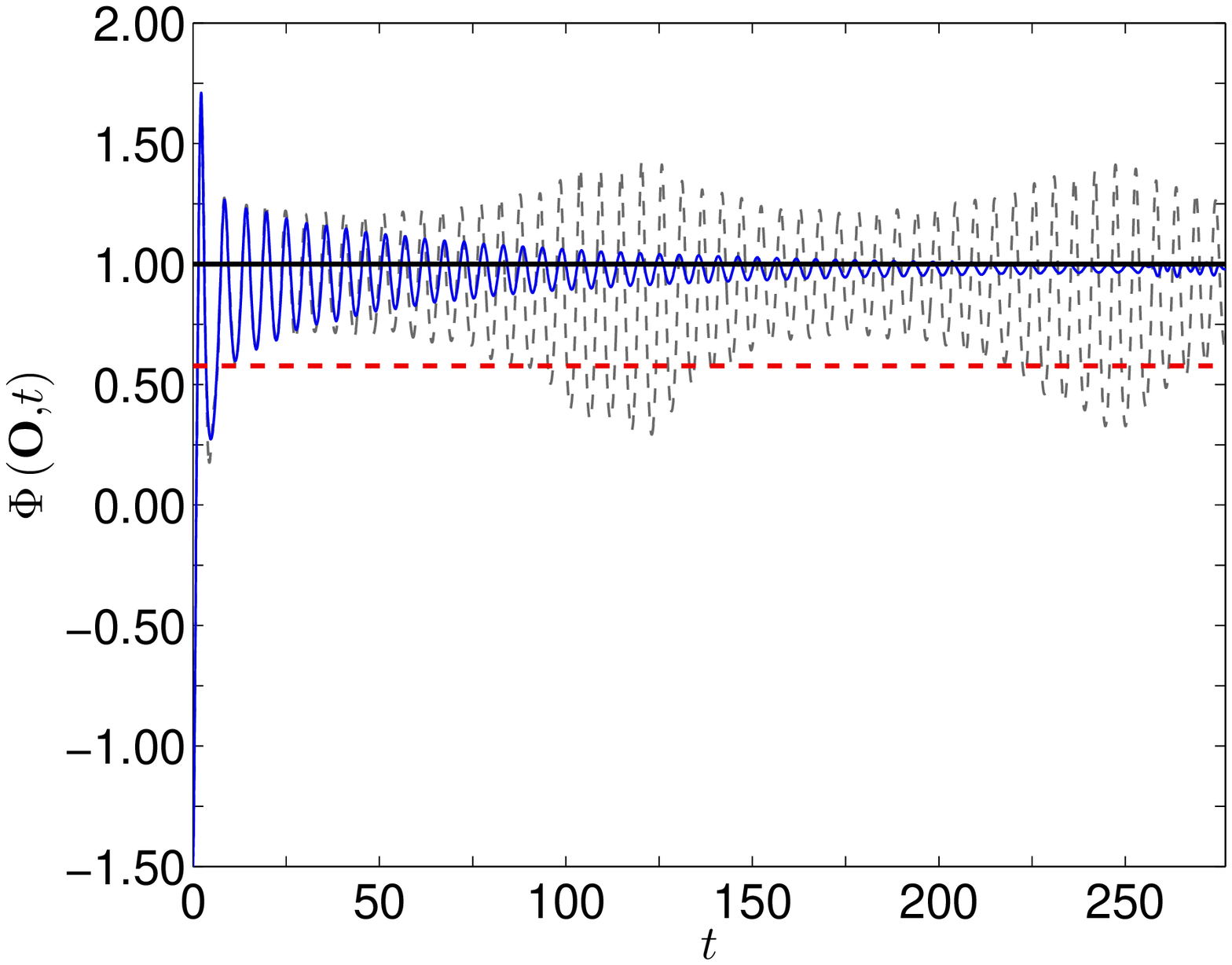}
    \includegraphics[width = 0.49\textwidth]{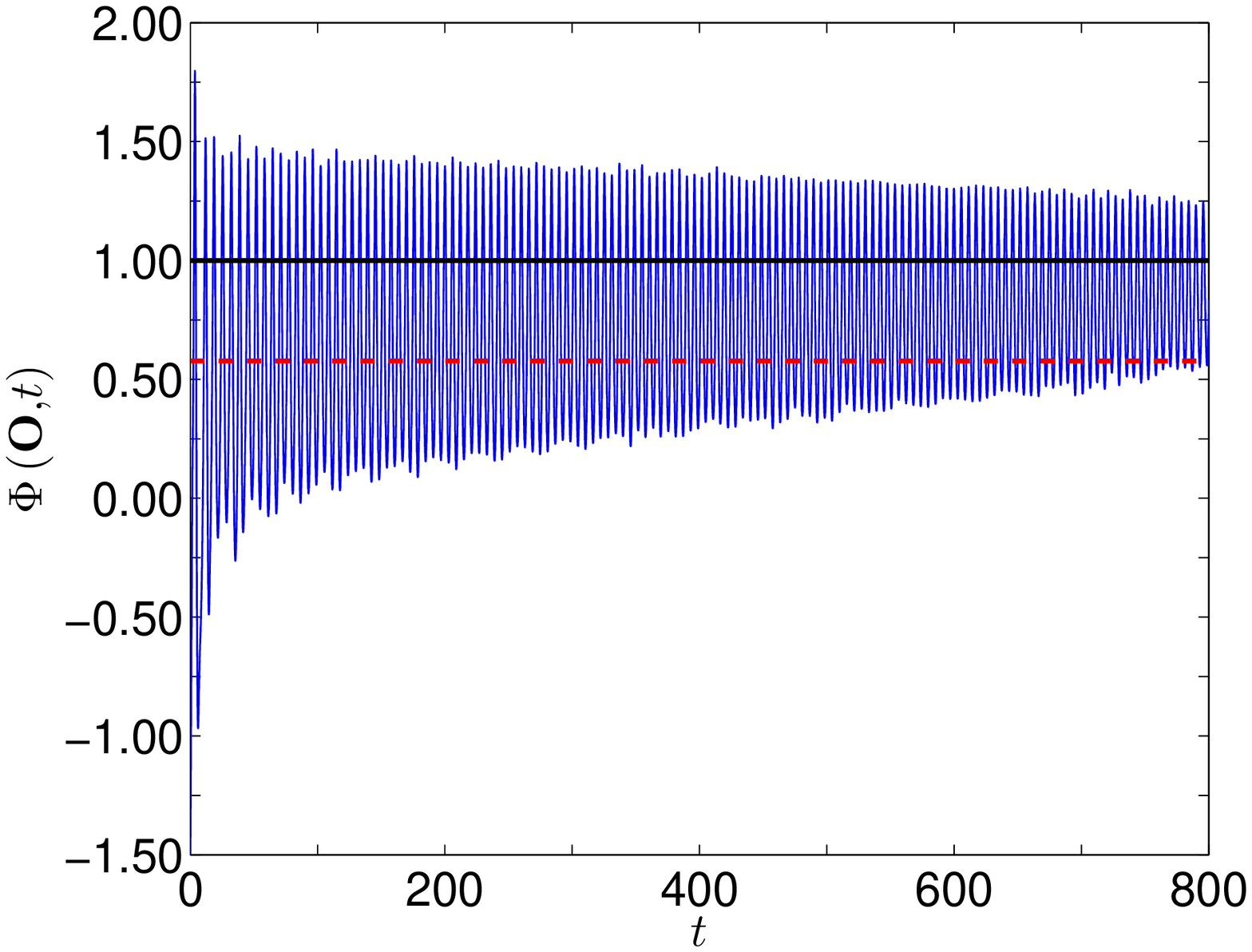}
  \caption{Initial conditions, whose evolution shows the unphysical small beat frequency phenomenon (left), and the physical large beat frequency phenomenon (right). $(A_0,r_0)=(2.5, 1.5)$ and $(2.5, 2.5)$, respectively. }
  \label{fig:small_large_beat}
\end{figure}

In some cases, simulations displayed an envelope around the basic oscillation, that was itself oscillating at a high frequency. Finer resolution simulations eliminated this superimposed oscillation (see Fig. \ref{fig:small_large_beat}, left).   These artefacts of the coarsest lattices entirely vanished on the increase in resolution for each initial profile examined, to leave only oscillations around the positive vacuum, with an amplitude below the point of inflection in the potential. This, essentially, leaves the initial profile in question outside the oscillon basin of attraction. Any distinctive appearance of comparably strong beats further to those explicitly examined were hence disregarded as unphysical and a signal of being outside the basin of attraction\footnote{These strong beats arose at disparate points in the whole parameter space of initial profiles examined.  The stark contrast in emergence of the beat frequency after a strong similarity in the oscillation to neighbouring points in parameter 
space, along with the 
highly peculiar nature of the beat frequency in comparison to the evolution on the oscillons in general was sufficient though to identify these beat frequencies, shown to be unphysical. Completing these larger simulations to confirm this in every such case was prohibitively time consuming.}. 

A beat frequency higher than in the evidently unphysical cases was observed for various other initial profiles, and the presence of this beat frequency in the cases tested was not removed in the repetition of selected simulations, at the higher spatial resolution (see Fig. \ref{fig:small_large_beat}, right). This suggestion of a physical beat frequency in the oscillons mimics the observations in the classical system \cite{bi:Hindmarsh2006}\footnote{Similarly, the study \cite{bi:Hindmarsh2006} highlighted that similar beat frequencies occur much more strongly in the case of the Sine-Gordon potential.}. The amplitude and the frequency of the modulations, however, alters on the change in the resolution of the lattice, perhaps indicating a non-physical component to these minor beat frequencies. Regardless, the continuation much longer than the natural timescale of the system and oscillatory characteristics on each change in resolution remained, characterizing the oscillon. The permanence confirmed the physical 
nature of the quantum oscillons, and any comparably weak beat frequency modulating an apparent oscillon further to those explicitly examined on the larger lattices is hence regarded to be physical.

\begin{figure}[H]
  \centering
    \includegraphics[width = 0.8\textwidth,angle=0]{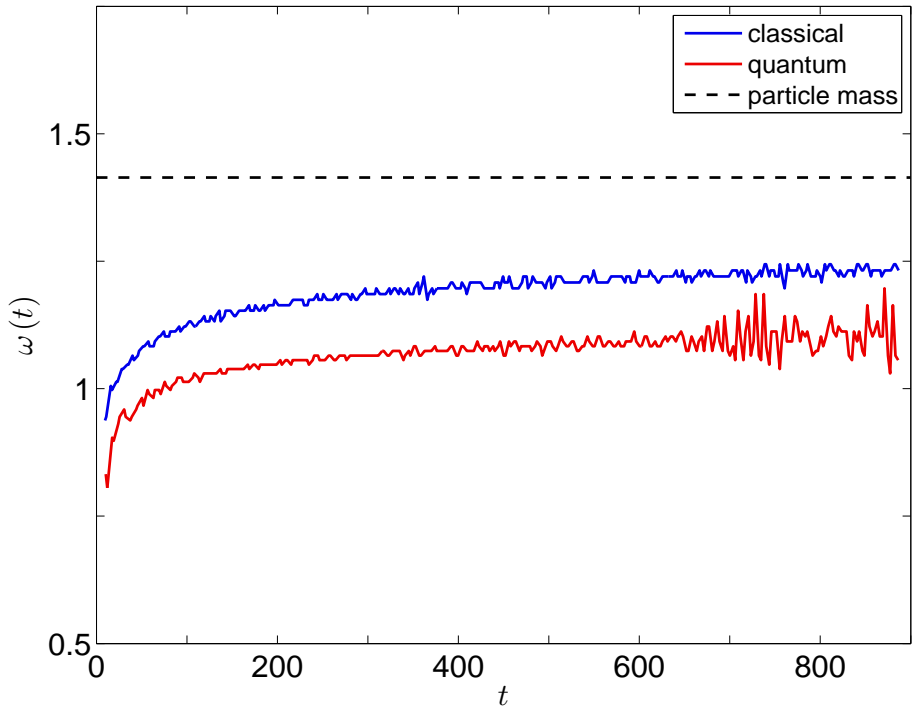}
  \caption{The frequency for a classical (blue) and the corresponding, quantum (red) oscillon. $(A_0,r_0)=(2.25, 2.5)$.  The black, dashed line shows the frequency corresponding to the particle mass in the broken phase $\sqrt{2}m$.}
  \label{fig:freq}
\end{figure}

We end this section by in Fig.~\ref{fig:freq}, showing the evolution of the oscillation frequency as a function of time for one particular initial condition in the classical (black) and the quantum (red) case. We see that for the classical case, the frequency settles at some value smaller than $\sqrt{2}m=1.41...$ (in units where the mass parameter is $m=1$), as is characteristic for an oscillon. The basic particle-like excitation has mass $\sqrt{2}m$. In the quantum case, the frequency is also lower than the particle mass, and even lower than the classical case. However, due to the quantum corrections, it is not clear what the natural frequency of the quantum oscillon should be. 
  Nonetheless, because of the renormalization, if the correlator is close to vacuum at the end of the simulation, the particle excitation in the quantum case should  be close to the classical value. The envelope passes the inflection point around $t=460m$, but we see in the frequency there that nothing dramatic happens. For a classically decaying oscillon the frequency would suddenly jump from distinctly below $\sqrt{2}m$ to $\sqrt{2}m$. For the quantum case, the frequency has a gradual evolution until at late times, around $t=600m$, it becomes less regular. This highlights the ambiguity in defining the lifetime of the quantum oscillon. We checked that the gradual decrease in central amplitude is not the result of the oscillon having moved away from the centre of the lattice.

\section{Classical and quantum basins of attraction}
\label{sec:basin}

A convenient (if not exact -- see previous section) way to define the lifetime of an oscillon is when the envelope of the oscillation (at the centre point of the oscillon) crosses the inflection point of the potential
\be
\frac{d^2V}{d\phi^2}=0\quad \rightarrow \quad \phi=\sqrt{\frac{m^2}{3\lambda}},
\ee
which for our parameters become $\phi=\sqrt{1/3}=0.577...$. Examples for the time evolution of an oscillon is shown in Fig.~\ref{fig:liveanddie}, for the classical (top left) and quantum (bottom left) case.  The solid line is the broken phase minimum, around which the oscillon evolves, and the dashed line is the inflection point. As mentioned, in the classical case, the oscillon settles but then continues to oscillate for the length of the simulation. The envelope never goes within the inflection point. For the quantum case, this happens around time $mt=460$ that we correspondingly define to be its lifetime. Clearly, the decay of the amplitude of oscillon is very gradual, and so the precise lifetime will depend on this definition. Nonetheless, because it is gradual, shifting the definition of the decay point does not influence the relative lifetime of oscillons evolved from different initial profile. 

We also note, based on the analysis above, that we understand the oscillon to have decayed (or be in the process of decaying) when a strong beat frequency appears in the evolution, even if the maximum of the modulation recrosses the inflection point. For the case of the small beat-frequencies, the defined endpoint may be when the net, modulated field first crosses below the point of inflection. The endpoint could also be based on the evolution of the envelope when this may be determined. As a consequence of this ambiguity, small-beat-frequency configuration have some systematic error in the lifetime determination. Any difference in the lifetime calculated from the envelope or the net, modulated field in each simulation however is minimal compared to the calculated lifetime.  Equally, the observed change in the frequency of the small modulations on switching to the larger lattices, according to either method adjusts the instant of the decay little compared to the corresponding, measured lifetime.  

\begin{figure}[H]
  \centering
  \includegraphics[width = 0.9\textwidth]{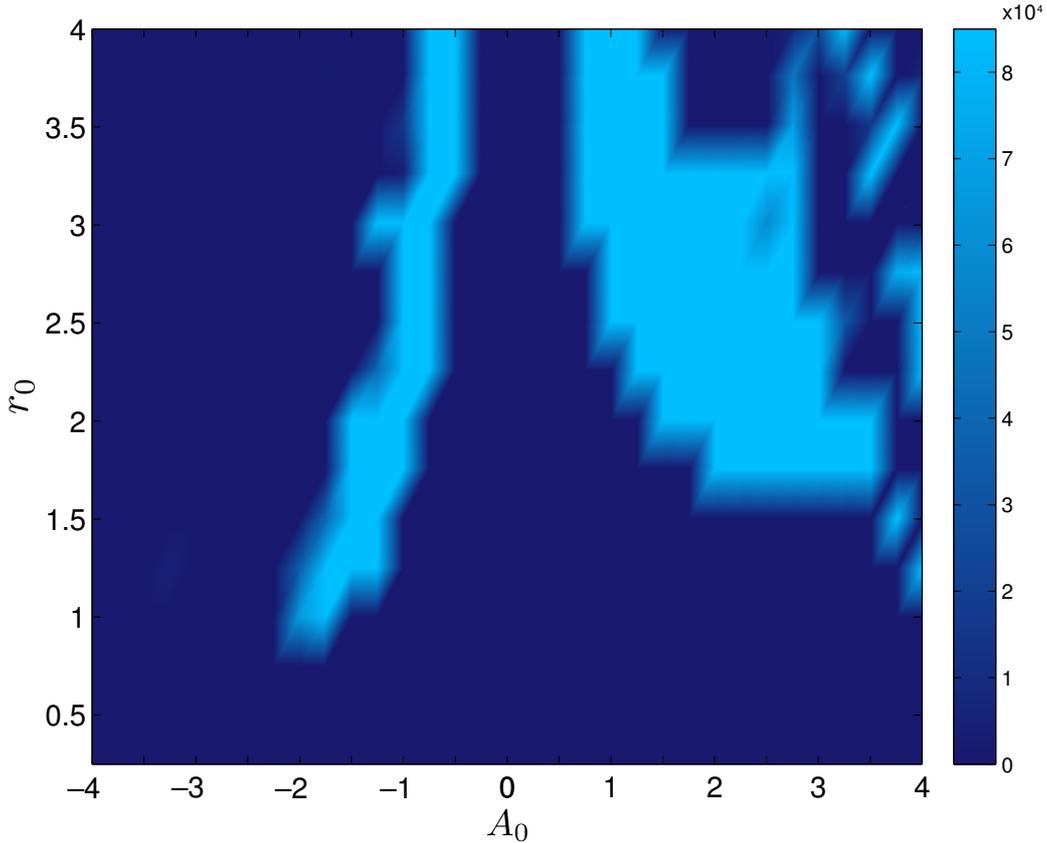}
  \caption{The lifetime of classical oscillons projected onto the parameter space of Gaussian initial conditions. }
  \label{fig:oscillon_lifetime_clas}
\end{figure}

We first consider the basin of attraction for classical simulations, i.e. ignoring the quantum modes and their back reaction on the mean (classical) field. This is numerically straightforward, and has been studied in some detail in \cite{bi:Andersen2012}. Fig.~\ref{fig:oscillon_lifetime_clas} shows the region of interest, as part of the whole space of initial configurations in the Gaussian parametrization.

We see that there are two main regions for positive and negative amplitude $A_0$, separated by a throat of instability, roughly corresponding to amplitudes less than the inflection point of the potential. There is also a lower limit to the width $r_0$, corresponding to the need for a certain size and energy to be "near", in the sense of the basin of attraction, to the oscillon. As was also demonstrated in \cite{bi:Andersen2012}, in two spatial-dimensions, classically stable oscillons are stable for a very long time. 

It is worth noting that the two-region structure is slightly misleading; they correspond to the two extrema of the oscillon oscillations and although the exact point-by-point matching is not obvious, each Gaussian released from rest in the right-hand region will approximately correspond to a Gaussian released from rest in the left-hand region. It is therefore sufficient for us to sweep in half the parameter space.

\begin{figure}[H]
  \centering
  \includegraphics[width = 0.9\textwidth]{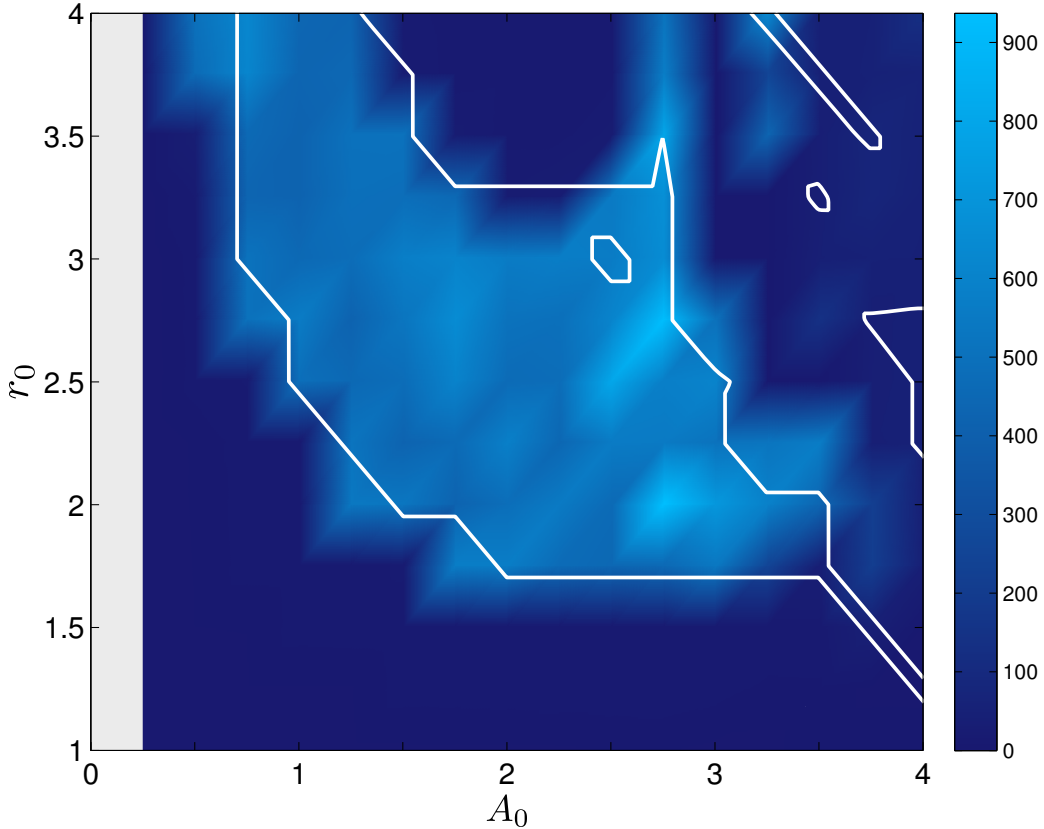}
  \caption{The lifetime of quantum oscillons projected onto the parameter space of Gaussian initial profiles. The white contours indicate the classical basin.}
  \label{fig:oscillon_lifetime}
\end{figure}

Including quantum fluctuations, we expect the effective potential to be different from the classical. Further, because the mean field is inhomogeneous and time-dependent, the effective potential will not simply be the usual perturbatively computable one but involve gradients in space and time. Moreover, because the mean field spans a significant fraction of the non-linear potential, it is not even clear that a gradient expansion would be sufficient.

In \cite{weir} a similar exercise was performed with Q-balls, where the field varies in time and space but the effective mass of the propagator (the modulus of the field) varies only in space and is constant in time. In that case, although notably in three spatial-dimensions, the effect of quantum fluctuations on the stability of the system was significant.

Fig.~\ref{fig:oscillon_lifetime} shows the equivalent of Fig.~\ref{fig:oscillon_lifetime_clas}, when including quantum fluctuations and their back reaction. We use the exact same initial profiles. Note that this is not a further approximation since also in the classical case, the Gaussian is equally only-approximately an oscillon; so what we map is whether different Gaussians are in the basin of attraction for the oscillon of the classical and/or the quantum system.

We see that the structure is very similar to the classical one. There is a sub-inflection throat region for small $A_0$, a minimum width $r_0$ and some additional structure relative to the overall main region. However, the actual lifetimes are much shorter, and near the edges of the main region, we have the physical (large frequency) and unphysical (small frequency) beating cases described earlier. We assign the small frequency cases to be outside the stability/oscillon region. 

\section{Conclusion}
\label{sec:conc}

We have performed the first quantum dynamical simulations of oscillons, using the inhomogeneous Hartree approximation. This is numerically very challenging since the mean field oscillon is spatially large, space- and time-dependent, and since in order to not have emitted energy travelling around the lattice and influencing the settling oscillon, we need a volume much larger than the oscillon itself. In addition, we found that replacing the full set of Gaussian quantum modes by an ensemble or random realizations did not in practice reduce the numerical effort. 

Our results show that at least for the parameters chosen here ($\hbar = \lambda/m = 1$), quantum fluctuations significantly shorten the lifetime and evolution of the oscillons. Decay is gradual rather than instantaneous, and the precise decay process is still uncertain (as it is for the classical case). 

We also mapped out and compared the basin of attraction of the classical and quantum oscillons, and perhaps surprisingly found that although the lifetimes are shorter, the region of Gaussian profiles that result in an oscillon are roughly the same. This suggests that at least at early times, the oscillon profile is similar, and that if the Gaussian profile has enough energy and is otherwise "spatially big enough", it does not feel the presence of the fluctuating modes until at later times. 

Regardless, because oscillons are time-dependent, real-time simulations are the only way to study their quantum properties. It is, for instance, not possible to do a Monte Carlo integration of the path integral.  The approximation for the full quantum dynamics employed here can in principle be systematically improved by including more diagrams beyond the LO in 1/N or Hartree approximation, but then expanding in modes is no longer an option. One could also consider stochastic quantization with a complex action, which has been partially successful for scalar fields \cite{stocquant}.  It is also possible that employing the classical-statistical approximation and simply averaging over an ensemble of random fluctuation realizations on top of a classical profile is a fruitful avenue. This remains to be seen.

\vspace{0.2cm}

\noindent
{\bf Acknowledgments:}  AT thanks David Weir for useful discussions and collaboration on related topics. PS and PT acknowledge support by STFC. AT is supported by the Villum Foundation. The numerical work was performed on the COSMOS/DIRAC supercomputing facility, funded by STFC and DBIS UK.

\appendix

\end{document}